# Blade: A Blockchain-supported Architecture for Decentralized Services

Extended Preprint Version 1.0

Sebastian Göndör , Hakan Yildiz , Martin Westerkamp , and Axel Küpper
Service-centric Networking
TU Berlin | Telekom Innovation Laboratories
Berlin, Germany
Email:{sebastian.goendoer | hakan.yildiz | westerkamp | axel.kuepper}@tu-berlin.de

*Abstract*—Decentralized services and applications provide a multitude of advantages for their users, such as improved privacy, control, and independence from third parties. Anyhow, decentralization comes at the cost of certain disadvantages, such as increased application complexity or communication overhead. This aggravates the development and deployment of decentralized services and applications. In this paper we present Blade, a software platform that aims to ease the effort of development, deployment, and administration of decentralized services by implementing reusable solutions for recurring challenges developers are facing when designing decentralized service architectures. This includes functionality for e.g. identity management, access control, request handling, verification of authenticity and integrity, discovery, or routing. Blade implements all this functionality in a Blade server instance, which can be deployed on a lightweight device, such as a NAS, Raspberry Pi, or router at home. This allows users without expert knowledge to run a Blade instance with already existing hardware with little overhead. Blade supports polyglot Blade modules that implement extended functionality, such as interfaces, frontends, and business logic of decentralized applications, e.g. a decentralized instant messaging service or an online social network. Based on the Oracle GraalVM, Blade modules can be implemented in a variety of programming languages and utilize the functionality provided by the Blade server instance. Blade modules are published in a Ethereum-based decentralized marketplace from where they can be installed directly via the Blade instances. For identity management and discovery of endpoints, Blade relays on blockchain-based smart contracts. Identity management and discovery is implemented via an Ethereum-based registry, allowing users to create and manage their identities in a self-sovereign manner without any central entity controlling the process. This way, Blade builds a decentralized service ecosystem that supports developers of decentralized applications and services.

## I. INTRODUCTION

The initial design of the Internet and the World Wide Web is based not only on mere distribution of hardware and resources, but also on the idea of decentralization of power and seamless interoperability [2], allowing all participants in the Web to communicate and interact with each other, regardless of possible failure of other nodes in the network [3]. Besides independence from central entities governing the network, decentralization introduces several other advantages, such as privacy or control over one's (personal) data [4]. In order to protect their user's privacy, improve fault tolerance, or resistance against legal action from organizations or governments, several types of web services and applications therefore adopt a decentralized architecture. One of the most prominent examples is the invention of Blockchain technology [5], which introduces a decentralized network of nodes that commonly maintain and advance a synchronized ledger in a distributed fashion.

Aside from the aforementioned advantages, decentralization also introduces a set of disadvantages and challenges, most importantly communication overhead, system complexity, security and reliability, synchronization of data, and development overhead [6] [7] [8]. While these issues can be easily solved and implemented with a central entity in place that governs and controls the process, decentralized systems need to implement means to solve these issues without being able to rely on a central trustworthy authority. Although several promising architectures and services exist that are built in a decentralized fashion, the most common type of architecture is a centralized one. We argue that one of the reasons for a lack of more decentralized services is the fact that the overhead for design, implementation, and governance is too high for many projects. We therefore see a need for a framework or platform for decentralized services that allows developers and users to implement and use them in an easy and uncomplicated fashion.

We identified a set of repeating challenges and questions that surface when designing and implementing a decentralized architecture. These challenges comprise, for example, decentralization of identity management (IDM) and discovery, verification of authenticity and integrity of data and messages, storage and replication of user data, as well as message formatting and routing.

In this work we present Blade, a **Bl**ockchain-supported **A**rchitecture for **De**centralized services. Blade is an extensible software platform for building decentralized services and web applications in an easy and uncomplicated fashion. Blade implements decentralized solutions for IDM, discovery, communication, access control, and communication protocols in form of a base server software, while Blade modules, being polyglot service components that implement service functionality, can be installed as plugins in order to extend the base functionality. Such Blade modules may then utilize and access the functionality provided by base instances using dedicated APIs. For ease of discovery, deployment, and use of the Blade modules, a decentralized marketplace allows developers to publish such Blade modules and users to retrieve, install, and run these modules to use the respective decentralized service applications. This way, Blade builds an interoperable, decentralized ecosystem of applications and services, which allows users to benefit from the advantages of decentralization while at the same time support the process of development and installation of these services. Therefore, the design of Blade is in line with the principles layed out in the *Contract for the Web* [9].

The contribution of Blade is threefold: First, we present Blade, an extensible software architecture for decentralized, federated services and the Blade communication protocol, which is used for communication between Blade servers. Second, we present a blockchain-based identity and discovery mechanism for self-sovereign identity management of users in the Blade ecosystem. Third, we present a decentralized, blockchain-based marketplace that allows developers to publish extensions to the Blade platform in form of polyglot software modules. While solutions for the individual parts existed before, Blade combines the concepts in a novel manner. In this work, we give an overview of the platform and its components.

This paper is organized as follows: First, Section II presents and discusses our previous work as well as similar approaches and other solutions proposed in the field. While several approaches exist that aim to provide a framework or architecture for decentralized services, to our knowledge Blade follows a unique approach. Section III then describes the general concept and design of Blade, followed by a description of the architectural components the system in Section IV. Section V describes the protocol and data formats used, while the prototypical implementation of the Blade platform is described in Section VI. Finally, Section VII concludes the paper.

## II. RELATED WORK

Blade taps into various topics of research, such as decentralized services, blockchain-based services, self-sovereign identity (SSI), or communication protocols. This section will highlight the most relevant topics.

Decentralization of service architectures first became popular in the late 1990's when peer-to-peer (p2p) allowed sharing of files, such as music or movies. Decentralization of such file sharing services circumvented the bottlenecks of content servers and at the same time removed the legally responsible central entity from the network [10]. Today, popular applications of decentralized services can e.g. be found in the domain of (decentralized) online social networks (D-OSN). One prominent example is Mastodon[1], which implements a distributed, federated network of mutually independent servers that provide a Twitter-like functionality. Users can select any available server to create and host their profile and data - or run their own instance. Communication between users and instances is implemented via the open protocol ActivityPub [11]. Anyhow, as with most federated services, user identities are still inflexible as they are controlled by the owner of the server they were created on. In case such a server is shut down or refuses to host a user's data and identity anymore, user identities are essentially lost. Similar problems exist in other federated architectures and protocols, such as XMPP [12] or Matrix [13].

The advent of Distributed Ledger Technology (DLT) or blockchain revived the trend towards a decentralized web [5]. To further decentralize service governance and data synchronization, blockchain systems, such as Bitcoin [5] or Ethereum [14], distribute a globally synchronized *state* across a large number of *nodes* called *miners* [15]. This global state is organized as an immutable ledger of information, to which changes can only be appended in form of transactions. Such a transaction is able to update a specific part of the ledger's *state*, depending on the permissions of the sending entity, where the mining nodes have to reach a *consensus* of which transactions are accepted [16]. To further extend the concept of blockchains to a "*world computer*" [14], *smart contracts* were introduced. Smart contracts can be deployed in such a ledger, acting as autonomous software agents that execute their logic when triggered by a transaction. Such smart contracts are able to store and perform operations on state variables that can be used to persist data in the ledger.

Besides implementing decentralized cryptocurrencies, blockchain technology is used for building decentralized applications (dApps), which implement all program logic either in client applications or in smart contracts [17]. Following this paradigm, dApps are ultimately independent from any central entity and solely rely on the distributed ledger to operate. Besides for dApps, smart contracts have been used for various other use cases. For example, smart contracts allow to store mappings of e.g. names to network locations and therefore facilitate a foundation for DLT-based registry systems, such as the DNS [18]. One example is the Ethereum Name Service (ENS)[2], which implements an alternative name registry in the Ethereum blockchain for the primary purpose of having human-readable identifiers for otherwise rather cryptic addresses. The system is based on a flexible concept of a central registry contract and flexible resolvers, which allow any registered name to be resolved to various different information, including Ethereum addresses, IP addresses, or other information.

---

[1]Mastodon: http://joinmastodon.org
[2]Ethereum Name Service: https://ens.domains

While smart contracts fundamentally allow building decentralized storage solutions on DLT, numerous reasons against storing data on-chain exist, cost of storage (transaction cost), write performance (transaction throughput), or privacy (public accessibility of data) being the most important ones. To decentrally store data, other systems may be used, for example the InterPlanetary File System (IPFS) [19], which is a p2p-based content-addressable storage system. The IPFS replicates data objects to multiple participating nodes using the Kademlia-based overlay *libp2p*[3], where individual objects are identified by a hash-based Content ID[4]. Anyhow, data objects stored in the IPFS are publicly accessible via their Content ID and cannot be deleted or overwritten. The IPFS is therefore not suitable for storing sensitive user-generated content, such as private messages or photos.

### A. Other Approaches

Blade utilizes a variety of technologies and results from research and development projects in the field of decentralized applications and services conducted in recent years.

Sonic is a project that proposes a decentralized and heterogeneous ecosystem of Online Social Network (OSN) platforms [20], [21]. End users of Sonic-based OSN services remain in full control of their social profiles, identities, and data by decoupling social profiles from the services they were created in. A decentralized Kademlia-based IDM service [22] links distributed identities of individual users to their respective social profile locations. Following the approach of Sonic, arbitrary OSN services implement a common RESTful API, which supports all standard functionality of state-of-the-art OSN services, while supporting extended and future functionality via protocol extensions. Users in the Sonic ecosystem are then able to connect to and communicate with other users using arbitrary other OSN services in the ecosystem, rendering platform gaps transparent and irrelevant. The project ReThink implements a communication architecture for peer-to-peer calling for Over-the-Top (OTT) services [23]. Following the ReThink approach, devices are discovered based on a distributed, public-key-based identifier, which is resolved to a device's network address via the Identity Mapping and Discovery Service (IMaDS) [24]. Once connected, devices of communication partners negotiate a communication protocol, for which a plugin is automatically discovered, downloaded, and installed for further communication. This way, communication partners may use different implementations of a service - or different services altogether - to seamlessly communicate with each other. Tawki [25] is a decentralized microblogging framework and application that allows users to setup and use their own communication infrastructure. Similar to other federated communication architectures such as XMPP [26], Mastodon [27], or Matrix [13], Tawki allows users to setup and run an individual server, which acts as an endpoint for other users in the system. Messages set to a user are directed to the target user's respective server, from which the addressee then can fetch them e.g. using a mobile client. Discovery and IDM functionality was implemented via a custom resolver in the ENS, allowing users to both register and control their identities in the Ethereum blockchain.

Besides the aforementioned approaches, the project Solid proposes a similar concept for decentralization of social web applications [28], [29]. In Solid[5], users install a personal data storage (POD) on their own web server, which is linked with a WebID-TLS identity. Solid PODs then implement RESTful interfaces for access to the managed data according to the Linked Data Platform (LDP) specification. This way, Solid enables developers to easily build applications for a decentralized, self-hosted environment. The main difference to Blade is that Solid assumes that all user data is made available through interoperable formats and accessible at a user's POD, while users control access to the data records via access control lists (ACL). This implies the existence of open and agreed-on data formats for everything handled by Solid apps and dictates that all apps comply with such data formats. While the semantic web hasn't yet succeeded to deliver its promise of a general machine-readable web of semantic information [30], extensive vocabularies for various use cases and application scenarios exist (e.g., Activity Streams 2.0 [31], *Schema.org*[6], or the Friend-of-a-Friend vocabulary (FOAF) [32]). Anyhow, services and applications rarely fully comply to these standards, which would cause incompatibilities in interoperable scenarios [33]. The approach proposed in this work addresses this issue through definition and ad-hoc negotiation of communication protocols to use for interaction between Blade servers, by which the data formats used are implied by the respective protocol specification. This way, the issue of different implementations and use of data formats can be mitigated through agreement on and support for a specific protocol used by both communication partners. Blockstack [34] also follows a similar approach to Solid or Blade by building a decentralized naming system built on DLT. Client implementations in Blockstack closely follow the idea of dApps, while private user data is persisted on the Gaia storage system [35]. Gaia uses existing cloud storage, e.g. Dropbox or Google Drive, to persist data objects and makes them available via regular CRUD operations, while allowing the user to define ACLs. Blockstack dApps then access data records in Gaia when required [36]. While both Blockstack and Solid follow a similar idea, they aim to provide a solution for building dApps and services, instead of offering a solution for shared use of an installed software platform for multiple service installations per user. Therefore, the solution proposed in this work follows a more fundamental approach for building an ecosystem for decentralized services.

---

[3]ProtocolLabs libp2p https://libp2p.io/
[4]ProtocolLabs: Multiformats https://multiformats.io/
[5]Solid: https://inrupt.com/solid/
[6]Schema.org: https://schema.org/

## III. CONCEPT AND DESIGN

Building decentralized services requires developers to find solutions for a list of problems that can easily be solved in a centralized architecture, but introduce certain challenges when to be implemented in a fully decentralized fashion. For example, issuing unique identifiers for users or data objects is a rather simple task when orchestrated by one entity, but requires thorough orchestration when implemented as a fully decentralized system. Other examples besides identification are resolving identifiers to a network location, proof of identity for users, verification of integrity and authenticity of data, data storage and replication, access control management, or communication between the different nodes. For each of these issues, decentralized solutions have been proposed and implemented as an alternative to existing centralized counterparts. For example, for identification of users and data, Ethereum utilizes (part of) the hash of a user's public key, which is referred to as an *address* [14]. As the matching private key remains secret and is only known to its owner, authenticity of data can be verified by anyone via the author's digital signature.

Anyhow, even though solutions exist for these individual issues of decentralizing services and applications, a holistic and usable solution does not exist yet. Developers hence have to research and select suitable technologies for each new decentralized service architecture, which then have to be implemented and integrated in a decentralized system or service. The resulting systems often turn out not to be reusable without adaptations and therefore increase the complexity of development. Our vision for Blade entails a decentralized, privacy-preserving, and open architecture that supports polyglot extensibility and interoperability for third party functionality. This would create a decentralized service ecosystem that focuses on self-sovereignty of its users [37].

### A. Blade Concept

With Blade, we envision a software platform for designing and implementing decentralized services and applications that readily implements all basic functionality for decentralization. This platform must be able to run multiple types of decentralized services simultaneously, using a single solution for core functionality, such as identity management, message routing, or security. Blade proposes that every user runs his own server on which all user data is stored and managed. The data is made available to other users via specific APIs that define who can access what information. Such a Blade server can be installed on any compatible device, such as a network attached storage device (NAS), internet router, Raspberry Pi, or (managed) web server. With a broad adoption of broadband internet connections at home and more performant hardware in routers and other networked appliances, hosting web application and services from home is a realistic option [38]. Individual servers implement a common RESTful API and federation protocol for server-to-server (S2S) communication. Following the approach of self-hosted services and applications therefore guarantees that every user remains in full control of his data as access to all data can be controlled and restricted easily.

The S2S federation protocol connecting the individual servers supports extensibility of its functionality. While the server itself only implements basic functionality, polyglot software modules may be installed that extend the scope of the protocol for specific use cases. For example, a module implementing a photo sharing service might implement a protocol extension that allows access to images, which can then be accessed by Blade servers with the same protocol extension installed. Modules therefore implement their own business logic and protocol extension, and may offer a graphical user interface (GUI) or API for client-to-server (C2S) communication. Such clients can be web apps or mobile applications a users employs in order to communicate with his own Blade server via the C2S interface. Following this approach, each user in the Blade ecosystem only directly communicate with their own server, while all communication with other users and Blade servers is relayed to the respective communication target's server (see Figure 1).

In order to identify and discover individual Blade servers, a decentralized IDM and registry service is used. Users in Blade are identified via Ethereum addresses, which are mapped to the respective server's network location. The mapping of user identifier to server address is stored in a smart contract that facilitates easy and reliable lookup of the information. To further streamline the use and administration of Blade servers in terms of accessibility and usability, Blade implements a novel concept of a decentralized marketplace, which allows users to publish, search and install Blade modules from; directly in the Blade server software, thus recreating the usability of state-of-the-art app stores of today's mobile operating systems.

### B. Blade Identity Management

Following the idea of a fully decentralized service ecosystem, Blade employs self-issued and user-controlled identities that are registered in a decentralized registry service in the Ethereum blockchain. Implemented as a smart contract in Ethereum, the Blade registry allows any user to create an elliptic curve (EC) key pair called *identity key pair* and register a new identity that can then only be managed by the respective identity's owner.

Identifiers in Blade are twofold: Firstly, the Ethereum address, i.e. the last 20 Bytes of the keccak256-hash of a user's public key [14], are used as the main identifier. Anyhow, such addresses are not human-readable and hard to memorize. Therefore, Blade additionally registers a unique username upon identity creation, which is tethered to the user's identity and can therefore be resolved to the associated identity information. This way, Blade implements a flexible self-sovereign identity concept that fulfills all three features (decentralization, security, meaningfulness) for decentralized identifiers specified by Zooko's triangle [39].

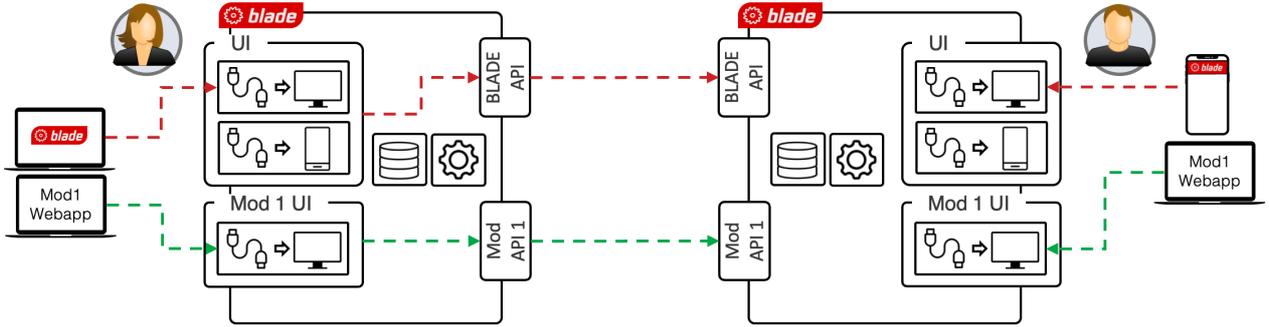

Fig. 1. All communication is relayed between individual Blade servers, while users only communicate with their respective own Blade server.

Besides registering an identity's address and unique name, Blade registers the network location of the Blade server associated with an identity. This allows anyone to resolve both a username as well as an identity's address to the associated network location, to which messages can then be directed. The identity owner may update the network location parameter to reflect a changed IP address or when moving one's Blade server instance to a new domain. This gives users the optimal flexibility in regards to their network location, for example in case of a frequently changing IP address at home.

As the security of identities in Blade is directly dependent on the secrecy of the identity key pair, a second cryptographic key pair is registered as a delegate for each identity. Such a delegate is created by computing a second key pair, the *delegate key pair*, which can be revoked and exchanged at any point in case it is considered compromised. This allows users to store the identity key pair's secret key at a secure location, such as a mobile wallet application (see Section VI-B), i.e. not on the Blade server itself. While delegate key pairs are used to sign content as well as requests and responses in the Blade ecosystem, the identity key pair is only used in the process of identity creation and when a delegate needs to be revoked. User identities can therefore further be resolved to the respective delegate, so that users can determine if a message from a Blade identity signed with a separate delegate key has actually been sent by the identity's owner. Blade identities and the associated information are published in the Blade registry in the Ethereum blockchain. The details of the Blade registry are described in Section IV-D.

### C. Use Cases

The following use cases have been defined for the Blade ecosystem to highlight the benefits and advantages of Blade over other approaches for decentralized service ecosystems. These use cases aim to provide a better understanding how Blade is used by end-users and developers. For all use cases, we consider three users, Alice (*A*), Bob (*B*), and Charlie (*C*). Each user maintains a Blade server *S(X)* for user *X* and accesses its functionality via a Blade client *Cl(X)*. Users are identified via an Ethereum address *ID(X)*, which is derived from the keypair $K_{ID}(X)$. Furthermore, each user has a delegate keypair $K_D(X)$ and a unique username *N(A)*, both of which are registered in the Blade registry. The use cases for Blade are also described in [40].

*1) Installation and Registration:* Alice wants to host her own Blade server on a device in her home network. She downloads the latest version of Blade and installs it on the device of her choice *S(A)*. Using the administration panel, she creates a local user account which she will use to login to her own Blade server instance to access its functionality. During the registration process, two cryptographic keypairs are automatically created: her identity key pair $K_{ID}(A)$ and a delegate key pair $K_D(A)$. In order to connect to and communicate with other users in the Blade ecosystem, Alice needs to register her identity in the Blade registry. Using the administration interface of her Blade server, she registers her identity with the Blade registry, which is created using her identity key pair $K_{ID}(A)$. In this process, her identifier *ID(A)* is registered in the registry and stored together with her delegate key pair as well as the URL or public IP address of *S(A)*. During the registration, Alice also registers a globally unique username *N(A)*, which allows resolving to both her identifier and associated information.

*2) Contact Management:* Alice wants to add her friend Bob to her contact list in Blade. She authenticates with her Blade server and uses the search functionality to look for users called Bob, which returns a list of several different users named Bob. Each user is described with a profile picture and a textual description, allowing Alice to select her friend from the list and send a connection request to him. The connection request is digitally signed using Alice's delegate key pair $K_D(A)$ and sent to the server location specified in the Blade registry for $K_{ID}(B)$. The connection request is received by *S(B)* and saved. The next time Bob logs in, he is notified about the new connection request from Alice. Once he accepts the request, a response message is sent back to Alice informing her about his decision. In order to allow each other access to specific resources and features on their Blade servers, both Alice and Bob configure access control lists (ACLs) that specify who can access which parts of an API and for what data.

*3) Communication:* Alice wants to communicate with her friend Bob using a instant messenger service she installed on *S(A)* in form of a Blade module. As Bob is already in her contact list, she sends a request to the Blade registry to resolve *ID(B)* to Bob's server's network location. Knowing Bob's server's network location, she sends a feature request message to *S(B)*, which yields a list of APIs that are currently supported by *S(B)*. If Alice's IM module uses an API that is also supported by Bob's Blade server, both can directly communicate via this common API. If this is not the case, Alice will be presented with a list of Blade modules available that support an API that is supported by both Blade servers - or by Bob's server only if there is no match. Alice can then install a suitable module from the Blade marketplace.

Once a compatible module is found and installed, Alice's is able to send a message to Bob. Her Blade server will assemble a request that is addressed to *ID(B)*, specifying *ID(A)* to be the sender of the message. The request is signed with her private key $K_D(A)_{private}$. Upon reception of such a message, *S(B)* will extract the sender's ID, i.e., *ID(A)* and attempt to resolve it via the Blade registry. If successful, the associated public key $K_D(A)$ is used to verify the request's authenticity and integrity before the message is shown to Bob.

*4) Blade Module Development:* Charlie is a developer and wants to implement a new decentralized social image blogging service in the sense of Instagram. Using the Blade framework, he can utilize the functionality already implemented by the Blade server software, such as identification, authentication, routing and verification of messages, or access control functionality. This significantly reduces the implementation workload for Charlie, who simply has to implement his project's business logic and graphical UI.

The finished project is compiled into a Blade application package (`*.bpk`) file, which is made available for download, e.g. via the IPFS, a web server, or a Blade server. Charlie proceeds to register his module in the Blade registry, assigning it an Ethereum address as its identifier. The data stored in the registry comprises information about the location of the Blade application package, versioning information, and used APIs.

Furthermore, Charlie registers the API his module implements in the Blade registry. Similar to the module, the API is assigned an Ethereum address as its identifier, which is resolvable to information about itself, such as versioning information and a link to a formal specification document that details specifics of the API. The module's registry record is updated to comprise the IDs of all APIs it implements. API and module IDs are published by Blade servers that run the respective modules/APIs, allowing other Blade servers to determine a suitable mode of communication.

## IV. ARCHITECTURE

The architecture of the Blade ecosystem comprises four main components, being Blade servers, Blade modules, the Blade registry, and Blade clients.

### A. Blade Servers

Following the idea of Blade, each user runs and maintains his own Blade server, on which all his data and information is stored. Blade servers can be deployed on managed or virtual servers, but may also installed on a user's networked appliances at home, such as a NAS or router. As routers and NAS appliances nowadays typically provide sufficient computing performance and storage capacities and are often operated in a 24/7 manner, such appliances provide an ideal platform for running a Blade server from home [38] [41] [8]. Using the Blade protocol (see Section V), other users may request or send data from or to a Blade server, where the network location of individual Blade servers is discovered via the blockchain-based Blade registry. As each Blade server is operated and configured by the respective owning user, users remain in full control of their data and how other users may access it. This ensures privacy and control and top level data protection.

Blade servers implement a base software for the Blade ecosystem in form of an application server. This software platform implements a list of basic functionality required for the operation of the Blade ecosystem. The most important functionalities provided by the Blade server platform are:

- **Blade API:** For communication with other Blade servers in the Blade ecosystem, servers implement an extensible API and functionality for request processing and handling. While the Blade server's base API supports only basic functionality, such as requesting a list of installed APIs and Blade modules, or requesting basic user profiles, it can be extended by installing service modules as described in Section IV-B, which add functionality to server instances. The request handling capabilities of Blade include functionality for building and dispatching well-formatted and digitally signed requests to other Blade servers as well as the routing and verification of authenticity and integrity of incoming request and response messages.
- **Identity management:** The decentralized user identities used in Blade are based on elliptic curve key pairs that are registered in the Ethereum blockchain. The identity management capabilities of Blade implement functionality to create and manage a user identity and furthermore allow resolving user names and identifiers of other users to the respective network location and delegate information as described in Section III-B.
- **Blade GUI:** For accessing the functionality of one's Blade server, each Blade server implements an administration interface in form of a web app. The owner of the respective Blade server instance can log in to this Blade GUI to configure and administrate the instance. Part of the Blade GUI is the frontend for the Blade marketplace (see Section IV-E) via which users can search for and download and install Blade modules into their Blade server.

- **Access control:** To allow users to define fine-granular access permissions for other users that attempt to access data or information provided by a Blade server or Blade module, Blade servers implement functionality to specify ACLs for different interfaces and data records. This allows the owner of a Blade server to individually specify who can access what part of his data in what way, for example allowing a specific user read access to images published as part of a Blade module.
- **Data storage:** Blade allows Blade modules to store and manage data records in a database, while at the same time ensuring that Blade modules cannot read from or write to data records of other Blade modules. All user data is therefore persisted in a local database on the Blade server of the data's owner and only accessible to other users via the Blade protocol. This way, users remain in full control of their data, as data is always stored directly on the user's own instance.
- **Module Marketplace:** To streamline the usability of using the Blade ecosystem, Blade implements a decentralized marketplace for Blade modules. The marketplace registers Blade Modules as well as the respective APIs and allows developers to publish their own implementations for other users to download and use. Downloaded Blade modules are installed and executed on Blade servers and by that extend the functionality of the Blade server instance.

### B. Blade Modules

While the Blade server instances implement base functionality for connectivity and identity management in the Blade ecosystem, service logic for decentralized services is implemented in form of Blade modules. Blade modules are software plug-ins that are executed in the Blade server environment, where the lifecycle of each module is controlled by the Blade server. For maximal flexibility for software developers, Blade modules are run using the Oracle GraalVM, a polygot virtual machine that allows performant execution of code written in various programming or scripting languages, including Java, JavaScript, PHP, Perl, Python, or C++. This gives developers maximal freedom and flexibility in terms of choice of technology and programming language.

For interaction with the Blade server instance, Blade servers implement an internal API Blade modules can access. This way, Blade modules can access the base functionality of Blade servers as described in Section IV-A in a controlled fashion, for example for dispatching requests, adding another user to the address book, or storing data to the local data storage of the Blade server.

Blade modules comprise three main architectural components as depicted in Figure 2. The first component is the Blade module's business logic, implementing the functionality of the decentralized service realized by the Blade module. Blade modules are highly versatile and can implement any functionality from simple static web pages to highly complex applications. One example for a Blade module could be a decentralized photo sharing service, where one can upload, organize, and even share photographs directly from one's Blade server, or a complex decentralized online social network. In such an example of a decentralized photo sharing service, the Blade module's business logic would implement the functionality to upload, store, list, like, delete, and comment on the published photos, locally or on another user's Blade server. Images and other data would be stored and managed by the respective creator's Blade server, being accessible for other users via a dedicated module API.

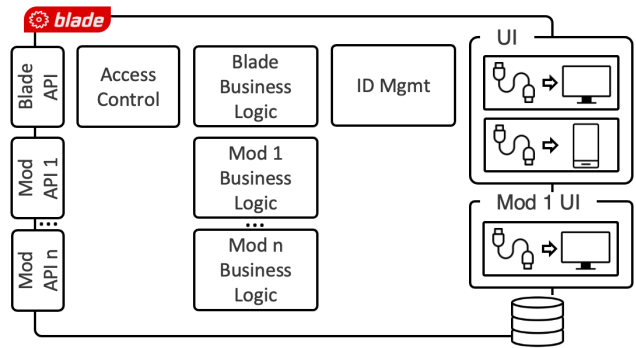

Fig. 2. Blade Server Architecture. Blade servers communicate via the Blade and Module APIs with other Blade servers (S2S), while users access the functionality via Blade clients via the Blade server and Module GUIs.

For communicating with other Blade servers, Blade modules implement a unique module interface that realizes the desired communication between the different instances of the module. This module API acts as the interface for communication with the same - and compatible - Blade modules that run on other Blade servers as depicted in Figure 3. For the example of the decentralized photo sharing service, the module API would support requesting a list of photos published on a Blade server, requesting individual photos, or posting a comment for one of the photos.

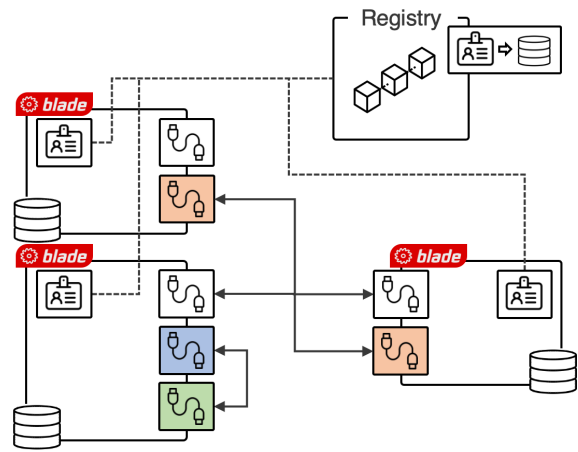

Fig. 3. Blade APIs: Service functionality of Blade servers is defined by the supported APIs. Blade servers can communicate with servers providing the same or compatible APIs. Furthermore, Blade modules can consume other module's APIs.

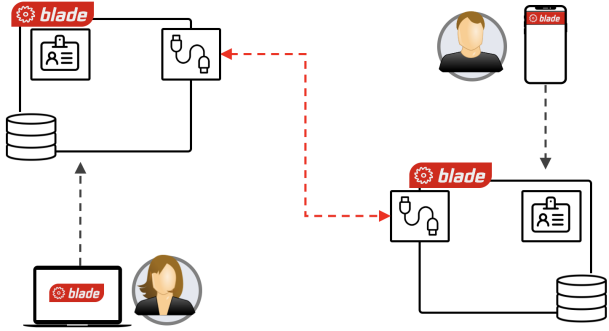

Fig. 4. Blade Communication: All communication is relayed between individual Blade servers, while users only communicate with their respective own Blade server.

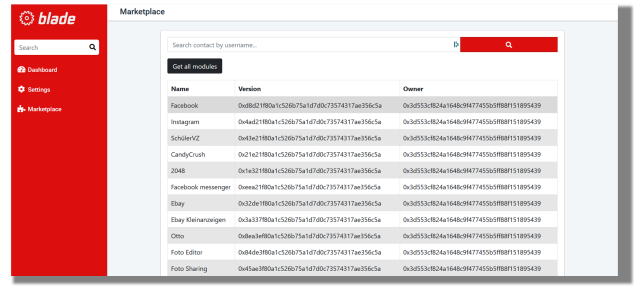

Fig. 5. Blade Marketplace: Blade modules and APIs can be searched for, downloaded, and installed directly via the Blade admin GUI.

In order to determine which module APIs are supported on a specific Blade server, the Blade API provides access to a list of supported APIs and their respective versions (see Section V). This allows users to determine the protocols and APIs used to communicate with a specific Blade server. Knowing which APIs are supported by the Blade server, users may install a Blade module with the same (or any compatible) API version on their Blade server in order to communicate. As described in Section IV-D, APIs need to be specified and registered as open source, allowing anyone to develop an application for any API - or specify new APIs or API versions altogether. This way, different Blade module implementations may be used, which are automatically detected as compatible and therefore communicate via their advertised APIs. This way, Blade fosters an open and heterogeneous ecosystem for decentralized services.

Upon installation of a Blade module, the new module is registered in the Blade server's internal module registry. The list of registered module APIs can then be accessed by other Blade servers that want to communicate. By disclosing which Module APIs are supported by a Blade server, other users can compile a list of compatible Blade modules available in the Blade marketplace. Installing such a compatible Blade module directly enables other users to connect and communicate. For example, if a Blade server implements a specific API for a messenger application, other users can install a compatible client supporting the same API in order to exchange messages.

### C. Blade Clients

In the Blade ecosystem, users rely on Blade client applications to access the functionality of their Blade servers. As depicted in Figure 4, Blade clients exclusively communicate with the Blade server of the respective user, which in turn relays all communication to other users' Blade servers. Blade clients can be implemented as mobile or desktop applications - or as web applications being served by a Blade server directly.

An example for a Blade client is a mobile application for an instant messaging application that allows a user to read and compose messages directed at other users in the Blade ecosystem. New messages composed by such an application would be sent to the respective user's Blade server, which would then forward them to the respective communication partner's Blade server.

### D. Blade Registry

The Blade registry implements a decentralized identity management service that allows users to create, register, and manage identities without relying on a central entity. As described in Section III-B, user identities are created and controlled by public key pairs, of which the associated Ethereum addresses are used as identifiers. Such identifiers are mapped to the respective user's Blade server location in the registry and can be resolved by any other user.

The Blade registry is designed as a smart contract in the Ethereum blockchain, allowing it to operate independently from central entities or organizations. This way, Blade implements a fully decentralized and independent identity management, creating an IDM with decentralized, secure, and human-readable identifiers [39]. Besides managing identities and resolving identifiers, Blade allows registration of organizations, which act as logical groups of users for publishing and managing Blade modules and APIs in the Blade marketplace (see Section IV-E).

### E. Blade Marketplace

The Blade marketplace is implemented as a part of the Blade registry, i.e. as a smart contract in the Ethereum blockchain. Similar to registering identities and organizations, the Blade marketplace implements functionality to register Blade modules as well as individual module APIs. Each registered Blade module and API is identified by an address, which is mapped to information about its author (i.e. organization), versioning information, and source location, from where the binary installation files (Blade modules) or the technical specification (Blade API) can be retrieved.

As Blade modules implement functionality and GUI of a decentralized service, they act similar to applications downloaded to and installed on a mobile phone from an app store. Users can access and search the list of registered Blade modules via the administration panel of their Blade servers and select individual modules to retrieve and install.

Blade servers publish a list of all APIs and their versions installed, which allows other users to determine a suitable Blade module to install in order to communicate with this specific Blade server. This allows a flexible way of installing software modules required for communicating with any Blade server in the ecosystem on the fly.

## V. BLADE PROTOCOL

For communication between individual Blade servers, Blade specifies an extensible RESTful API and S2S protocol that implements CRUD-operations. All requests are addressed to the identity address of the recipient, which is automatically resolved via the Blade registry to the current API endpoint of the recipient's Blade server. This way, Blade servers relay all communication in a loosely-coupled fashion directly to the targeted server, which then verifies and processes the request.

The Blade protocol is a RESTful application layer protocol based on HTTP/HTTPS for stateless message exchange between individual Blade server instances. Requests and responses specify a set of mandatory HTTP headers that comprise the identity addresses of both sender and receiver of a message, additional protocol metadata, and a digital signature of the entire request. This allows the recipient of such a request or response to verify whether the received message has actually been sent by the stated author, guaranteeing authenticity, non-repudiation, and integrity. The list of HTTP headers used is specified in Table I.

The payload of both requests and responses is encoded as a signed JSON Web Token (JWT) [42] [43] using the respective sender's delegate key pair for digitally signing the encoded payload. The message payload is specified by a private claim `data` in the JWT body, while the data schema of the payload itself is dictated by the respective Blade module used and is validated upon reception by the business logic of the receiving module. A request sent in the Blade ecosystem always addresses a target identifier. The target's identifier is first resolved to the associated URL of the recipient and the associated public key of the recipient's communication key pair via the Blade Registry as described in Section IV-D.

Blade servers publish a base API, which implements basic accessibility, such as retrieving information about the server instance, installed modules, or requesting basic information about the owner. In addition, each Blade module installed on a Blade server specifies its own module API, therefore extending the API. All module APIs must be registered in the Blade marketplace as described in Section IV-E, rendering each module API and API version uniquely identifiable via an id. This allows Blade servers to easily determine which compatible APIs are implemented by any given Blade server by querying the base API `GET /interfaces`. Requests directed to a specific Blade modul, i.e. API, are then sent to the endpoint `/:apiID/*`.

A typical scenario of communication in the Blade ecosystem between Alice and Bob looks as depicted in Figure 6: A priori, Alice needs to know the ID or username of the owner of Bob, which is then resolved via the Blade registry to the server's network location (Step 1). When now Alice wants to send a text message to Bob using a specific module, she first requests a list of installed API versions from Bob's server. This request is digitally signed and comprises the header information as specified in Table I (Step 2). Bob's server will receive the request and will verify its authenticity and integrity, which involves verifying Alice's identity via the Blade registry (Step 3) before sending back the requested information (Step 4). From the returned list, Alice determines a subset of API versions that are compatible with the APIs implemented by the module to be used. If this list is empty, Bob's server does not support the type of communication used. Otherwise, the message is encoded for a compatible API version and sent to Bob's server (Step 5). Bob's server will receive the request and will verify its authenticity and integrity, encoding any response to be sent back to Alice in a similar fashion.

TABLE I
BLADE REQUEST/RESPONSE HEADERS

| Header | Description |
|---|---|
| `BladeSourceID` | Identifier of the sender. |
| `BladeTargetID` | Identifier of the recipient. |
| `BladeProtocolVersion` | Protocol version. |
| `BladeNonce` | Random value. |
| `BladeSignature` | ECDSA signature created by the sender. |

## VI. IMPLEMENTATION

The prototypical implementation of Blade comprises four components: The Blade server instance, a Blade mobile wallet application, the Blade registry, and a Blade module demonstrator. This section elaborates on the details of the prototypical implementation. The code will be made available on GitHub once considered mature[7].

### A. Blade Server Prototype

The Blade server has been implemented as a prototype based on the Oracle GraalVM 22.1[8] using Java 17 and Spring Boot 2.6. Using the support for polyglot projects via the GraalVM, Blade modules can be developed using a broad variety of programming or scripting languages, giving developers the flexibility to choose a suitable language and framework. The Blade server implements two main interfaces, the administration GUI, which is realized as a ReactJS web app using AdminLTE 3.1[9] (C2S interface) and the Blade REST API for communication with other Blade servers (S2S interface). The administration GUI is depicted in Figure 7. Internally, the Blade server implements functionality for identity management, request handling, access control management, and polyglot module support.

---
[7]Blade at GitHub: https://github.com/blade-net
[8]Oracle GraalVM: https://www.graalvm.org/

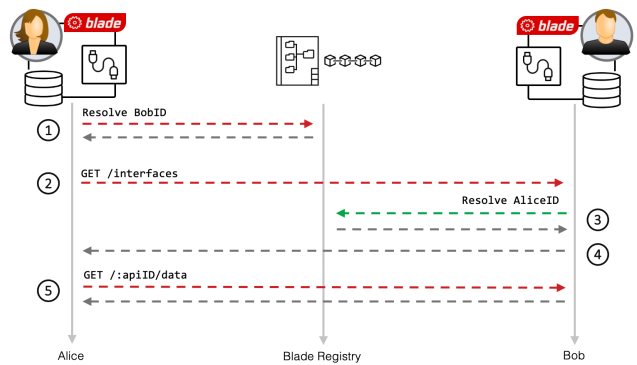

Fig. 6. Overview of the Blade communication flow. User identifiers and names are resolved via the Blade registry, while requests are addressed directly to the Blade server base API or available Blade module APIs.

The identity management allows users to create and manage their own identity, which is controlled by an EC key pair created by the user. On initial setup of an identity, the Blade server registers the identity in the Blade registry along with the required information and delegate keys as described in Section IV-D. Changes to the Blade server's network location are then automatically written to the Blade registry, facilitating seamless availability also in cases of frequently changing IP addresses, for example for dialup connections at home. Furthermore, the prototype implementation provides functionality for the Blade marketplace, from which Blade modules can be installed, access control management for incoming requests for other users, and basic contact management.

### B. Blade Wallet Application

In a scenario in which a user's Blade server is compromised, a key pair stored on and managed by the same server would be compromised, too. The consequence in such a case would be that the identity could not be trusted anymore and needs to be revoked. To prevent the identity key pair's private key from being compromised in such a scenario, identities in Blade may be created and managed by an external wallet application. The Blade wallet application is implemented in Java using Android 8 or higher (cf. Figure 8) and allows users to create and manage Blade identities directly on their phones. This way, the identity private key does not need to be stored on the Blade server anymore and can be securely kept in the wallet application. Therefore, in case the Blade server is compromised, the associated Blade identity remains secure.

The wallet application implements all functionality for creating and managing key pairs for a Blade identity as well as associated delegate keys, synchronizing changes to the Blade registry, and manage a contact address book of other known and trusted Blade identities. In case the user exchanges the current delegate key or edits his contact list via the wallet application, the changes are automatically synchronized to both Blade registry in the Ethereum blockchain as well as to the Blade server. This allows a more secure and user friendly management of Blade identities for users.

### C. Blade Registry

The Blade registry implements the IDM and marketplace functionality of the Blade ecosystem and has therefore been implemented as a smart contract for the Ethereum blockchain. The Solidity-based implementation[10] is loosely based on the ERC1056 standard [44] and has been deployed in the Ropsten testnet at `0x8708975b585762a09aa568736a5298d6845772b7`. The smart contract manages identities in mapping structures, where the addresses associated with the public key pairs of users are stored. Similar to identity mappings in ERC1056, the Blade registry registers an address as an identifier, where the matching key pair is used to control the identity by setting and updating parameters. Creating an identity in Blade registers the delegate, current network location of the associated Blade server, and a name mapping[11] for an identity, which can later on only be updated by the owner of the identity himself. While registration and management of an identity requires a transaction that accrues costs denoted in Gas [14], resolving a name or an identifier to the respective stored parameters does not cost any Gas and is therefore free of charge. Furthermore, all changes to identities, organizations, and the marketplace are emitted by the Ethereum Virtual Machine (EVM) as events, allowing clients to listen to changes in the registry.

---

[9]AdminLTE: https://adminlte.io/

[10]Blade Registry on GitHub: https://github.com/blade-net/blade-registry

[11]The name mapping in Blade is assigned in a first-come-first-serve basis during the registration process of a new identity and cannot be changed after.

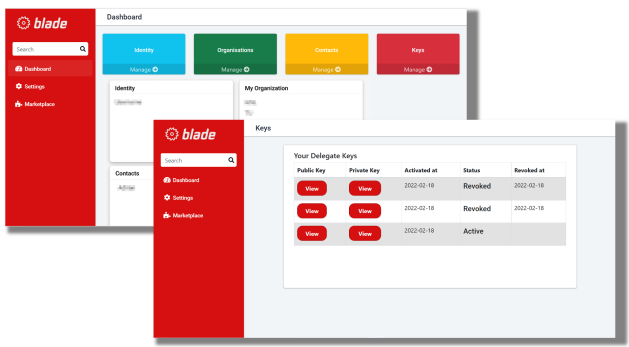

Fig. 7. Blade administration GUI and mobile Blade wallet application for identity management.

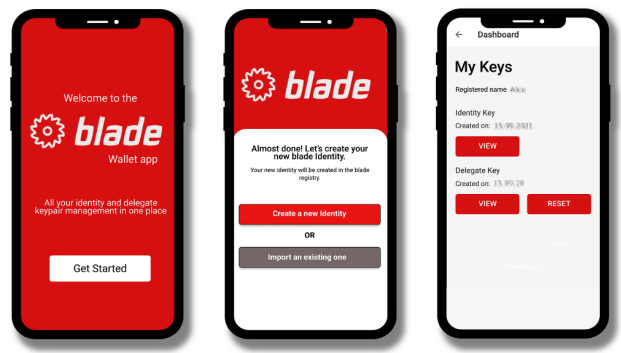

Fig. 8. Blade wallet application UI for mobile identity management.

TABLE II
BLADE REGISTRY GAS COSTS

| Function | Gas cost | Function | Gas cost |
|---|---|---|---|
| `createIdentity()` | 144,406 | `createOrganization()` | 120,779 |
| `setURL()` | 33,101 | `addOrgMember()` | 48,810 |
| `setDelegate()` | 55,481 | `removeOrgMember()` | 26,888 |
| | | `changeOrgOwner()` | 30,221 |

To estimate the cost of registration and managing identities in the Blade registry, the execution costs have been analyzed. While creating an identity is rather cost intensive with 144,406 Gas, updating an identity's network location is affordable with 33,101 Gas. Setting a new delegate address and revoking the old one in the process is similarly cheap with 55,481 Gas. Similarly, creating an organization costs 120,779 Gas, adding a new member to an organization costs 48,810 Gas, removing a member costs 26,888 Gas, and changing the owner of an organization costs 30,221 Gas. The results of this analysis are summarized in Table II. As of May 12, 2022, the Gas price for a transaction to be mined within five minutes was at 30 GWei, according ETH Gas Station[12]. Following this calculation, creation of a new identity amounts to approx. 0.00433 Ether. Even though the initial Gas costs are lower than those calculated for our previous work in [25] (144k vs. 207k Gas), through significantly higher Gas and Ether prices the resulting cost for registering a new identity in the Blade registry is comparatively expensive with approx. 8.73 USD. This can be explained by the highly fluctuations in the valuation of Ether and Gas costs. Anyhow, the evaluation shows that identities can be registered in the Ethereum main network for a rather small fee.

### D. Blade Module Demonstrator

Serving as a prototype and demonstrator for the applicability of the concept, a Blade module realizing a decentralized contact database has been implemented. Based on the idea of our previous work in [45], this Blade module allows users to host their contact data within the Blade module on their own server. The module enables users to subscribe to other user's contact data, which is hosted on the respective other user's Blade servers. If any user updates his contact information, changes are pushed automatically via the module's APIs to the Blade servers of all subscribers, ensuring that all users have the most recent version of that data record.

The Blade module further implements a RESTful C2S-API, which is consumed by a mobile Android app. This app automatically synchronizes the contact details received by the Blade server with the contact database on the user's phone, thus automating the process of updating the phone's contact database. To protect the privacy of users, the server implementation allows users to specify a per-user and per-item access control list, facilitating selective disclosure of contact information. The S2S API implements management for subscriptions to other users via `/:apiID/subscription[/:sID]` and pushing updated contact details to subscribed users via `/:apiID/subscription/:contactID`. This way, a decentralized, self-updating, and privacy-aware contact management is implemented.

## VII. CONCLUSION

In this paper, we presented Blade, a **Bl**ockchain-supported **A**rchitecture for **De**centralization. Following the idea of a fully decentralized and distributed ecosystem for self-hosted services, Blade facilitates users to run their own instances of fully decentralized services on a single integrating platform. This way, users remain in full control of storage and processing of as well as access to their data. While Blade servers implement base functionality of the platform, third-party applications may be installed in form of polyglot modules. These modules then implement functionality and user interfaces of a decentralized service, such as a decentralized messenger application. In order to streamline the process of discovery and installation of Blade modules for users, Blade implements a decentralized marketplace, via which users can search for available implementations and download and install them via the web-based GUI of the Blade server.

For identification management and discovery, Blade implements an Ethereum-based registry service, following which all users are identified by their Ethereum address. The Blade registry maps these addresses to the network locations of the users' Blade servers, so that identifiers are resolvable to the respective user information. When communicating with other users' Blade servers, Blade relies on an HTTP-based protocol that allows users to seamlessly communicate with each other. The protocol allows verification of authenticity and integrity of messages by digitally signing content and requests with the respective sender's keypair. While the Blade server's API only implements basic functionality, the protocol can be extended by Blade modules, thus allowing a flexible extensibility of the functionality. This way, Blade presents a foundation for creating an open and heterogeneous ecosystem for decentralized services and applications.

---

[12]ETH Gas Station: https://ethgasstation.info/